\documentclass{elsart}
\usepackage{amsfonts}
\usepackage{amsmath}
\usepackage{amssymb}
\usepackage{graphicx}

\begin{document}
\begin{frontmatter}

\title{Natural braneworld inflation and baryogenesis}

\author{R. Gonz\'{a}lez Felipe}
\address{Departamento de F\'{\i}sica and Centro de F\'{\i}sica Te\'{o}rica de Part\'{\i}culas\\
Instituto Superior T\'{e}cnico, Av. Rovisco Pais, 1049-001 Lisboa, Portugal}

\begin{abstract}

In natural inflation models, the inflaton is a pseudo Nambu-Goldstone boson and
the flatness of the potential is protected by shift symmetries. In this
framework, a successful inflation requires the global symmetry to be
spontaneously broken at a scale close to the Planck mass. Such a high value of
the spontaneous breaking scale may not be legitimate in an effective field
theory. On the other hand, if natural inflation occurs during the
nonconventional high-energy era in braneworld cosmology, the conditions for the
inflaton slow rolling can be eased and the spontaneous breaking scale can be
lowered to values below the Planck scale. We examine the observational
constraints on this scenario and study the possibility that the baryon
asymmetry of the universe is generated by the decay of the pseudo
Nambu-Goldstone boson during the reheating era.

\vspace{5mm} \noindent \emph{PACS}: 98.80.Cq, 98.80.Es, 04.50.+h
\end{abstract}

\end{frontmatter}

\section{Introduction}
\label{intro}

Today there is a compelling evidence that the early universe underwent a period
of cosmological inflation~\cite{Guth:1980zm,Lyth:1998xn}. Inflation not only
elegantly solves several cosmological issues, including the horizon and
flatness problems, but it can also provide the initial conditions required for
the structure formation in the universe. In particular, the observations by the
Cosmic Background Explorer (COBE) and, more recently, the Wilkinson Microwave
Anisotropy Probe (WMAP) satellites~\cite{Bennett:2003bz} confirm the
predictions of a flat universe with a nearly scale-invariant spectrum of
adiabatic perturbations.

Despite the existence of several cosmologically viable inflaton potentials, the
construction of a ``naturally" flat potential is a difficult task from the
particle physics viewpoint. In the simplest chaotic inflationary scenarios,
inflaton field values above the four-dimensional Planck mass, $\phi \gtrsim
M_{P}$, are typically required to allow for a sufficiently long period of
inflation. Thus, one expects nonrenormalizable quantum corrections of the order
of $\mathcal{O}\,[(\phi /M_{P})^{n}]$ to destroy the flatness of the potential.
Moreover, the inflaton must couple to matter fields to efficiently reheat the
universe after the inflationary era. Such couplings could also destabilize the
potential. Among the few known candidates which preserve the scalar potential
nearly flat and protected against radiative corrections, supersymmetry has
undoubtedly received most of the attention so far. However, it has recently
been argued~\cite{Arkani-Hamed:2003mz} that this symmetry alone cannot
naturally provide a satisfactory model of inflation, if supergravity effects
are taken into account. Another natural candidate for the inflaton is provided
by non-linearly realized symmetries, such as those which involve a pseudo
Nambu-Goldstone boson (PNGB)~\cite{Freese:1990rb} or extra components of gauge
fields propagating in extra
dimensions~\cite{Arkani-Hamed:2003wu,Kaplan:2003aj,Arkani-Hamed:2003mz}.

In the simplest variant of natural inflation~\cite{Freese:1990rb}, the role of
the inflaton is played by a PNGB, and the flatness of the potential is
protected by a shift symmetry under $\phi \rightarrow \phi + {\rm constant}$,
which remains after the global symmetry is spontaneously broken. An explicit
breaking of the shift symmetry leads typically to a potential of the form
\begin{equation}\label{pot}
    V(\phi)= \mathit{\Lambda}^4\, [1 - \cos(\phi/f)]\,,
\end{equation}
where $\phi$ is the canonically normalized field, $f$ is the spontaneous
breaking scale and $\mathit{\Lambda}$ is the scale at which the soft explicit
breaking takes place\footnote{Note that the flatness of the PNGB potential is
natural in the 't Hooft sense~\cite{tHooft}: in the limit $\mathit{\Lambda}
\rightarrow 0$ the shift symmetry is restored.}. For large values of $f$, the
potential can be flat. However, in this framework the slow-roll requirements
set the bound $f \gtrsim M_P$. Such high values could be problematic and
difficult to justify in an effective field theory~\cite{Arkani-Hamed:2003mz}.
Moreover, one can expect the quantum gravity effects to explicitly break the
global symmetry. A possible solution to the above problems is to consider
higher-dimensional cosmological models, where our four-dimensional world is
viewed as a three-brane embedded in a higher-dimensional bulk\footnote{For an
alternative solution, based on the presence of two axions with a potential
provided by two anomalous gauge groups, see Ref.~\cite{Kim:2004rp}.}. For
instance, in the 5D version of natural inflation considered in
Ref.~\cite{Arkani-Hamed:2003wu}, the inflaton is the extra component of a gauge
field, which propagates in the bulk, and the flatness of its potential, coming
only from nonlocal effects, is not spoiled by higher-dimensional operators or
quantum gravity effects.

In this paper we shall consider another possible scenario for extra-natural
inflation. We shall assume that the inflaton is described by a standard
four-dimensional PNGB, with a potential given by Eq.~(\ref{pot}), but inflation
occurs during the nonconventional high-energy regime of the
theory~\cite{Bento:2001hu}. In braneworld cosmology, modifications to the
Friedmann equation~\cite{Binetruy:1999ut} (due to the linear dependence of the
expansion rate $H$ on the energy density $\rho$ at early times) can ease the
slow-roll conditions and enable inflation to take place at field values far
below $M_{P}$~\cite{Maartens:1999hf,Papantonopoulos:2004bm}, thus avoiding the
well-known difficulties with higher-order nonrenormalizable terms. Since in
this framework the observational data does not put direct limits on the
symmetry breaking scales, but rather on the ratios $f/M_5$ and
$\mathit{\Lambda}/M_5$, it is possible to obtain lower values of the natural
inflation scale for low values of the five-dimensional fundamental Planck mass
$M_5$. The latter is only constrained by demanding the brane terms in the
Friedmann equation to play a negligible role at the big bang nucleosynthesis
scale $\sim \mathcal{O}({\rm MeV})$.

After the inflationary epoch, the cold inflaton-dominated universe undergoes a
reheating phase, during which the inflaton oscillates about the minimum of its
potential, giving rise to particle and entropy production. In addition to
entropy creation, the right abundance of baryons must also be created. It is
then interesting to ask whether the inflaton field could also solve another
outstanding cosmological puzzle, namely, the explanation of the baryon
asymmetry of the universe. A particularly attractive and viable mechanism is
the so-called spontaneous baryogenesis~\cite{Cohen:1987vi}, which can naturally
occur in the early universe, if the $CPT$ symmetry is violated. In this
framework, the PNGB field associated with the broken baryon number plays a
crucial role. If the PNGB is derivatively coupled to a baryon current $J_B^\mu$
with an effective Lagrangian of the form
\begin{equation}\label{Leff}
    \mathcal{L}_{{\rm eff}}=\frac{1}{f}\,\partial_\mu \phi J_B^\mu\,,
\end{equation}
a net baryon asymmetry can be produced either in a Hubble-damped regime or in a
regime characterized by the decays of the inflaton as it oscillates about its
minimum~\cite{Cohen:1987vi}. In the latter case, and assuming that the PNGB
field couples to fermions which carry baryon number, the resulting asymmetry
turns out to be proportional to $\mathit{\Gamma} \phi_i^3/f$, where $\phi_i$ is
the inflaton field value at the onset of reheating and $\mathit{\Gamma}$ is the
inflaton decay width~\cite{Dolgov:1994zq,Dolgov:1996qq}.

By combining the constraints coming from natural inflation with the ones
inferred for the present baryon-to-entropy ratio, it is then possible to put
limits on the reheating temperature and the relevant scales required for
natural inflation and baryogenesis to be successfully realized in the above
braneworld scenario.

\section{Natural braneworld inflation}

In a cosmological braneworld scenario, where the space-time on the brane is
described by a flat Friedmann-Robertson-Walker metric, the Friedmann equation
receives an additional term quadratic in the energy density\footnote{We assume
that inflation rapidly dilutes any dark radiation term and that the
four-dimensional cosmological constant is  negligible.}~\cite{Binetruy:1999ut},
\begin{equation}
\label{Fried} H^2 = \frac{8\pi}{3 M_P^2} ~ \rho ~ \left(1 +
\frac{\rho}{2\lambda} \right) ~,
\end{equation}
where $\lambda$ is the brane tension, which is related to the fundamental
five-dimensional Planck scale $M_5$ through the equation
\begin{equation}
\lambda = \frac{3}{4 \pi} \frac{M_5^6}{M_P^2}~. \label{eq:MP}
\end{equation}
At sufficiently low energies, \mbox{$\rho \ll \lambda$}, the Friedmann equation
of standard cosmology is recovered. For nucleosynthesis to take place
successfully, the change in the expansion rate due to the $\rho^2$-term in the
Friedmann equation (\ref{Fried}) must be sufficiently small at scales $\sim
\mathcal{O}$(MeV). This leads to the lower bound $ \lambda \gtrsim
(1~\mbox{MeV})^4$. A more stringent bound, $\lambda \gtrsim (1~\mbox{TeV})^4$,
can be obtained by requiring the theory to reduce to Newtonian gravity on
scales larger than 1 mm.

In the slow-roll approximation and at high energies ($V \gg \lambda$), the
total number of $e$-folds during inflation is given by~\cite{Maartens:1999hf}
\begin{equation}
\label{efolds} N \simeq - \frac{4\pi}{M_P^2\lambda} \int^{\phi_{F}}_{\phi_{I}}
\frac{V^2}{V'} d\phi ~,
\end{equation}
where $\phi_{I}$ and $\phi_{F}$ are the values of the scalar field at the
beginning and at the end of the expansion, respectively. The value $\phi_F$ can
be computed from the condition ${\rm max}\{\epsilon(\phi_F),|\eta(\phi_F)|\}=
1$, where $\epsilon$ and $\eta$ are the slow-roll parameters, given in the
high-energy approximation by
\begin{equation}
\epsilon \simeq  \frac{M_P^2\lambda}{4\pi} ~
    \frac{V'^2}{V^3} \,, \quad
\eta \simeq  \frac{M_P^2\lambda}{4\pi} ~ \frac{V''}{V^2}~.
\end{equation}
The spectra of scalar \cite{Maartens:1999hf} and tensor
\cite{Langlois:2000ns,Huey:2001ae} perturbations at the Hubble radius crossing
are also key parameters during inflation. Their expressions at high energies
read as
\begin{equation}
\label{scalamp} A_s^2 =   \frac{64 \pi}{75 M_P^6 \lambda^3} ~
\frac{V^6}{V'^2} ~, \quad A_t^2  =  \frac{8V^3}{25 M_P^4 \lambda^2}
~.
\end{equation}
The scale dependence of the scalar perturbations is described by the spectral
tilt
\begin{equation}
n_s-1 = \frac{d \ln A_{s}^2}{d \ln k}\simeq -6 \epsilon + 2 \eta~,
\end{equation}
and its running is given by
\begin{equation}
\alpha_s = \frac{d n_s}{d \ln k}\simeq 16 \epsilon \eta -18 \epsilon^2 -
2\xi\,,
\end{equation}
where
\begin{equation}
\xi \simeq \frac{M_{P}^4\lambda^2}{(4\pi)^2}\frac{V' V'''}{V^4} \,.
\label{eq:xi}
\end{equation}
Finally, the tensor power spectrum amplitude can be parameterized by the
tensor-to-scalar ratio
\begin{equation} \label{eq:rs}
r_s = 16 ~ \frac{A_{t}^2}{A_{s}^2} = 24\, \epsilon \,.
\end{equation}
We also remark that the tensor spectral index $n_t$ is not an independent
parameter since it is related to $r_s$ by the inflationary consistency
condition $n_t=-r_s/8$, which holds independently of the brane tension
$\lambda$~\cite{Huey:2001ae}.

In natural inflation, with the potential given as in Eq.~(\ref{pot}),  the
following inflationary constraints are obtained from
Eqs.~(\ref{efolds})-(\ref{eq:rs}):
\begin{align} \label{allconstr}
N_* &= \frac{1}{x}\left[(\sin^2\!\chi_F - \sin^2\!\chi_*) + \ln
\left(\frac{\cos^2\!
\chi_F}{\cos^2\!\chi_*}\right)\right]\,, \nonumber\\
A_s &=\left(\frac{2}{75 \pi^2 x^3}\right)^{1/2}
\left(\frac{\mathit{\Lambda}}{f}\right)^2 \frac{\sin^5\! \chi_*}{\cos\!
\chi_*}\,, \nonumber\\
n_s &= 1 - \frac{x}{\sin^4\!\chi_*} \left(1+4
\cos^2\!\chi_*\right)\,,\\
\alpha_s &= -\frac{2 x^2\cos^2\!\chi_*}{\sin^8\!\chi_*} \left(3+
2\cos^2\!\chi_*\right)\,,\nonumber\\
r_s &= 24 x \frac{\cos^2\!\chi_*}{\sin^4\!\chi_*}\,,\nonumber
\end{align}
where $N_*$ is the number of $e$-folds before the end of inflation, at which
observable perturbations are generated, and
\begin{equation}\label{defx}
    \chi = \frac{\phi}{2f}\,,\quad x = \frac{\lambda}{8 \pi}
    \left(\frac{M_P}{f\mathit{\Lambda}^2}\right)^2\,.
\end{equation}
For a given value of the parameter $x$, the value of the field at the end of
inflation can be determined from the condition $\epsilon(\chi_F)=1$, leading to
the equation
\begin{equation}
   2 \sin^2\!\chi_F = -x+\sqrt{x^2+4x}\,.
\end{equation}
For $x \ll 1$, we obtain $\sin\!\chi_F \simeq x^{1/4} (1-\sqrt{x}/4)$.

\begin{figure}[t]
\begin{center}
\includegraphics[scale=0.7]{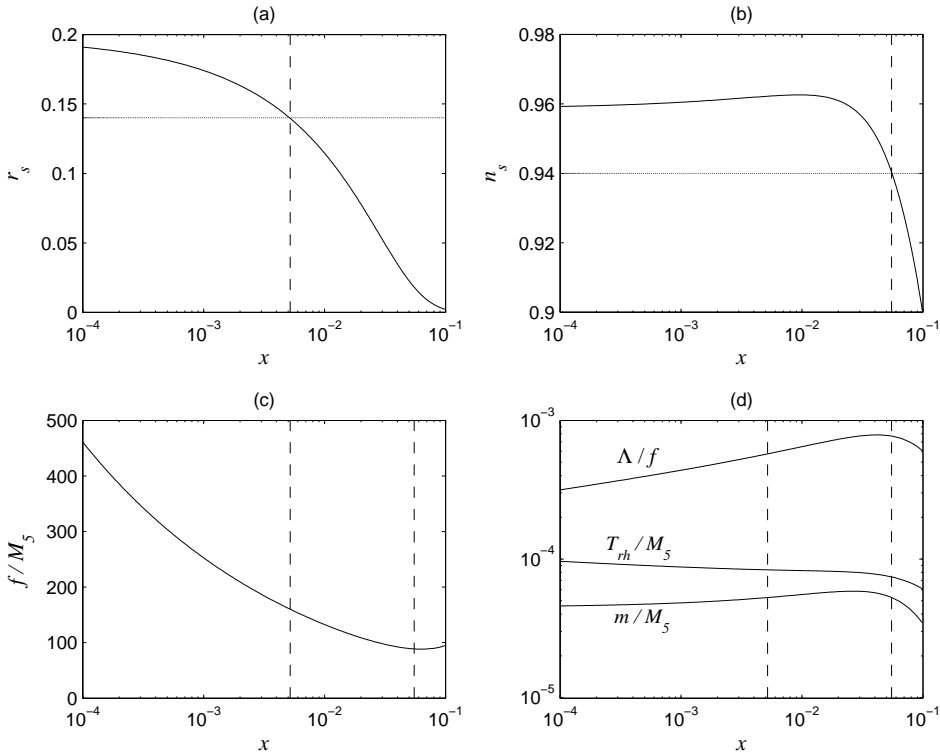}
\caption{Inflationary parameters $r_s,n_s$ (Figs. 1a and 1b), the symmetry
breaking scales $f, \mathit{\Lambda}$ (Figs. 1c and 1d) and the inflaton mass
$m = \mathit{\Lambda}^2/f$ (Fig. 1d) in the natural braneworld inflationary
scenario. The plots are given as functions of the dimensionless parameter $x$
defined in Eq.~(\ref{defx}). The region delimited by the vertical dashed lines
corresponds to the parameter space allowed by the WMAP bounds on $n_s$ and
$r_s$.} \label{fig1}
\end{center}
\end{figure}

In our numerical estimates we shall use the COBE normalization $A_s \simeq 2
\times 10^{-5}$ and the following WMAP bounds on the inflationary
parameters~\cite{Peiris:2003ff}
\begin{equation}
0.94 \leq n_s \leq 1.00~,~|\alpha_s| \leq 0.02~,~r_s \leq 0.14~,
\label{eq:WMAPConst2}
\end{equation}
which apply to models of inflation with a negative curvature ($\eta<0$). We
also recall that in standard cosmology a reasonable fiducial value for the
number of $e$-folds is $N_* = 55$. This number is expected to be higher in the
brane scenario~\cite{Dodelson:2003vq,Wang:2003qr}. In what follows we assume
$N_*=60$. It is worth noticing that for the potential (\ref{pot}) the spectral
index $n_s$ is very weakly dependent on $N_*$ and its running $\alpha_s$ is
negligible, $\alpha_s \approx -5 \times 10^{-4}$. Therefore, if small-scale
microwave background observations indicate a significant negative running of
the spectral index, natural braneworld inflation in its simplest version of
Eq.~(\ref{pot}) will be excluded.

The numerical solution of Eqs.~(\ref{allconstr}) is presented in
Figure~\ref{fig1}. We see that natural inflation in the brane constrains the
parameter $x$ to the range $5.2 \times 10^{-3} \lesssim x \lesssim 5.5 \times
10^{-2}$. The lower bound comes from the WMAP upper bound on $r_s\,$, while the
upper bound comes from the minimum allowed value for $n_s\,$. This leads to the
following constraints on the spontaneous and explicit symmetry breaking scales,
\begin{equation} \label{fconstraint}
90 \lesssim \frac{f}{M_5} \lesssim 160~, \quad 7 \times 10^{-2} \lesssim
\frac{\mathit{\Lambda}}{M_5} \lesssim 9 \times 10^{-2}~.
\end{equation}
which imply the hierarchy of scales $6 \times 10^{-4} \lesssim
\mathit{\Lambda}/f \lesssim 8 \times 10^{-4}$ and an inflaton mass $5 \times
10^{-5} M_5\lesssim m=\mathit{\Lambda}^2/f \lesssim 6 \times 10^{-5} M_5\,$. It
is also worth noticing that in the present framework the density fluctuation
spectrum is red tilted ($0.94 \lesssim n_s \lesssim 0.96$) and a significant
amount of gravitational waves is predicted ($ 0.02 \lesssim r_s \lesssim
0.14$), which could be detectable at forthcoming experiments such as the PLANCK
satellite.

We remark that the above analysis has been done assuming that the high-energy
approximation is valid, i.e., $ V/\mathit{\Lambda} \gg 1$, so that inflation
occurs in the nonconventional braneworld era. This requires $M_5 \ll
10^{17}~\mbox{GeV}$. On the other hand, in the range of $x$ allowed by
inflation, one obtains $\chi \sim \mathcal{O}(1)$. Therefore, from
Eq.~(\ref{fconstraint}) it follows that the inflaton field $\phi \sim f \sim
\mathcal{O}(10^2)\times M_5 \ll M_P$ in the high energy regime, i.e. natural
inflation indeed takes place at field values below the Planck scale.

\section{Reheating and baryogenesis}

After inflation, the cold inflaton-dominated universe undergoes a phase of
reheating, during which the inflaton decays into ordinary particles and the
universe becomes radiation dominated. Assuming an instantaneous conversion of
the inflaton energy into radiation, one can identify $\rho=\rho_{\rm
rad}=(\pi^2/30)\,g_\ast\, T^4$, where $g_*$ is the effective number of
relativistic degrees of freedom; $g_* \sim 100$ in the standard model for
temperatures above the electroweak scale. The reheating temperature $T_{rh}$ is
then obtained from the requirement that the expansion rate of the universe
equals the inflaton decay width, i.e. $H(T_{rh})=\mathit{\Gamma}$. In the
braneworld scenario, this leads to the relation
\begin{equation}\label{trh}
T_{rh}^4= \frac{T_t^4}{2}  \left[-1+\sqrt{1+\frac{45}{\pi^3 g_*} \left(
\frac{\mathit{\Gamma} M_P}{T_t^2}\right)^2}\,\right]\,,
\end{equation}
where $T_t$ is the transition temperature from brane cosmology to standard
cosmology, defined through the relation $\rho(T_t) = 2 \lambda\,$ so that
\begin{equation}\label{ttrans}
    T_t^2 = \frac{2}{\pi} \sqrt{\frac{15 \lambda}{g_*}} =
     \frac{3}{\pi} \sqrt{\frac{5}{\pi g_*}} \frac{M_5^3}{M_P}\,.
\end{equation}

During the oscillating phase, the inflaton field evolves according to the
well-known equation of a damped harmonic oscillator,
$\ddot{\phi}+\mathit{\Gamma} \dot{\phi}+m^2 \phi = 0$. If the inflaton field
couples derivatively to the baryon current through the interaction Lagrangian
of Eq.~(\ref{Leff}), the net baryon number density produced during reheating
can be estimated as~\cite{Dolgov:1996qq}
\begin{equation}\label{nB}
    n_B \equiv n_b - n_{\bar{b}} = \frac{\mathit{\Gamma}\phi^3_i}{2f} \,,
\end{equation}
where $\phi_i$ is the value of the field at the onset of the reheating epoch.

When the expansion is negligible ($H < \mathit{\Gamma}$), the baryon number can
be obtained from Eq.~(\ref{nB}) by replacing $\phi_i$ by the field value at
$T_{rh}$,
\begin{equation}\label{phirh}
    \phi_{rh} = \pi\sqrt{\frac{g_*}{15}}\, \frac{T_{rh}^2}{m}\,.
\end{equation}
Since the entropy density after thermalization is given by $s=2\pi^2 g_*
T_{rh}^3/45\,$, the baryon-to-entropy ratio reads then as
\begin{equation}\label{nBovers1}
    \frac{n_B}{s}=\frac{\pi}{4}\sqrt{\frac{3g_*}{5}}\, \frac{\mathit{\Gamma}}{f}\,
    \left(\frac{T_{rh}}{m}\right)^3.
\end{equation}
On the other hand, the process of particle production starts at earlier times,
when $H \simeq m> \mathit{\Gamma}$. Since the generation of the asymmetry is
more efficient at that time, the final baryon-to-entropy ratio is larger than
that of Eq.~(\ref{nBovers1}) approximately by a factor
\begin{equation}\label{dilution}
    \kappa = \mathcal{D}\,\left(\frac{\phi_i}{\phi_{rh}}\right)^3,
\end{equation}
where $\phi_i = \phi(H\simeq m)$ and
\begin{equation}\label{Dfactor}
    \mathcal{D} = \left(\frac{\mathit{\Gamma}}{m}\right)^2\,
    \frac{1+m \,M_P/\sqrt{3\pi \lambda}}{1+\mathit{\Gamma}\,M_P /\sqrt{3\pi \lambda}}
\end{equation}
is a dilution factor due to the expansion of the universe from the time $t \sim
1/m$ to $t \sim 1/\mathit{\Gamma}$. The final baryon-to-entropy ratio is
therefore given by
\begin{equation}\label{nBovers2}
    \frac{n_B}{s}=\frac{\pi\kappa}{4}\sqrt{\frac{3g_*}{5}}\, \frac{\mathit{\Gamma}}{f}\,
     \left(\frac{T_{rh}}{m}\right)^3.
\end{equation}
Let us note that $\kappa \simeq m/\mathit{\Gamma}$ in standard
cosmology~\cite{Dolgov:1996qq}, while in the high-energy regime of brane
cosmology we obtain $\kappa \simeq (m/\mathit{\Gamma})^{1/2}$.

Using Eqs.~(\ref{trh}) and (\ref{ttrans}), the decay width can be written in
the form
\begin{equation}\label{Gdecay}
    \mathit{\Gamma} = \frac{2\pi}{3}\sqrt{\frac{\pi g_*}{5}}\, \frac{T_{rh}^2}{M_P}
    \left[1+\left(\frac{T_{rh}}{T_t}\right)^4\right]^{1/2}.
\end{equation}
Moreover, the field ratio $\phi_i/\phi_{rh}$ will be given by
\begin{equation}\label{phiratio}
    \frac{\phi_i}{\phi_{rh}} =
    \frac{1}{\sqrt{2}}\left(\frac{T_t}{T_{rh}}\right)^2 \left[-1+
    \sqrt{1+\frac{45}{\pi^3 g_*}\left(\frac{m M_P}{
    T_t^2}\right)^2}\right]^{1/2}.
\end{equation}
Thus, in the high-energy regime of brane cosmology, i.e. when $\rho \gg
2\lambda\,$, one obtains from Eq.~(\ref{nBovers2}),
\begin{equation}\label{nBBC}
    \frac{n_B}{s}= \frac{\pi^2 g_*}{10}\sqrt{\frac{\pi}{6}}
    \frac{T_{rh}^5}{f m^{5/2} M_5^{3/2}}\,,
\end{equation}
while in standard cosmology,
\begin{equation}\label{nBSC}
    \frac{n_B}{s}= \frac{\pi}{4}\sqrt{\frac{3 g_*}{5}}
    \frac{T_{rh}^3}{f m^2}\,.
\end{equation}

\begin{figure}[t]
\begin{center}
\includegraphics[scale=0.6]{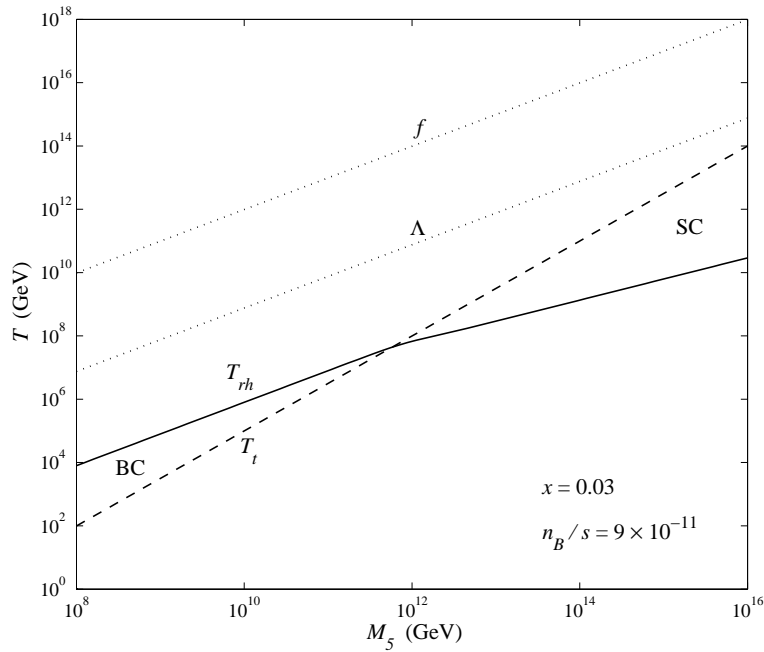}
\caption{The reheating temperature $T_{rh}$ (as a function of the 5D Planck
mass $M_5$) required for a successful baryogenesis in the natural braneworld
inflationary scenario. The dashed line corresponds to the transition
temperature from brane cosmology (BC) to standard cosmology (SC). For
comparison, the symmetry breaking scales $f$ and $\mathit{\Lambda}$ required
for inflation are also shown (dotted lines).}\label{fig2}
\end{center}
\end{figure}

The recent observations by WMAP~\cite{Bennett:2003bz} imply that $n_B/s \simeq
9 \times 10^{-11}$, a value which is in remarkable agreement with the
independent determination of this quantity from big bang nucleosynthesis. Using
the above value, we can put bounds on the reheating temperature required for
baryogenesis to take place during the radiation era. The results are presented
in Figure~\ref{fig2}, where $T_{rh}$ is plotted as a function of the
fundamental 5D Planck mass $M_5$. We notice that $T_{rh}$ is quite insensitive
to the variations of the dimensionless parameter $x$ defined in
Eq.~(\ref{defx}), in the range allowed by natural inflation (cf.
Figure~\ref{fig1}). In Figure~\ref{fig2} we present the results for $x=0.03$,
assuming $g_* = 100$. From the figure we see that baryogenesis through PNGB
decays can occur in the high-energy regime of brane cosmology for $M_5 \lesssim
6 \times 10^{11}$~GeV and $T_{rh} \lesssim 4 \times 10^7$~GeV, at much lower
reheating temperatures than in standard cosmology.

One may wonder whether the baryon asymmetry produced by the mechanism described
above will survive without being diluted by baryon number violating processes
in thermal equilibrium. In fact, if nonzero baryon and lepton numbers are
generated such that $B-L \neq 0$, the asymmetry will not be washed out by the
anomalous electroweak sphaleron processes. On the other hand, if an equal
amount of baryon and lepton asymmetries is generated such that $B-L$ vanishes,
the final asymmetry will be diluted by such processes, unless the reheating
temperature is below the critical temperature of the electroweak phase
transition $T_\mathrm{ew} \sim 100$~GeV. At higher temperatures, i.e. before
the electroweak symmetry breaking, the sphaleron rate is roughly $
\mathit{\Gamma}_{\mathrm{ws}}\simeq 25\, \alpha_{\mathrm{w}}^{5}\,
T$~\cite{Bodeker:1999zt}, where $\alpha_{\mathrm{w}} \simeq 1/30$ is the weak
coupling constant. Thus, electroweak sphalerons in standard cosmology are in
thermal equilibrium ($\mathit{\Gamma}_{\mathrm{ws}} \gtrsim H)$ at temperatures
$T_\mathrm{ew} \lesssim T \lesssim 10^{12}$~GeV. In brane cosmology, this bound
depends on the fundamental scale $M_5$ and we find $T_\mathrm{ew} \lesssim T
\lesssim 10^{-2}M_{5}\,$, in the high-energy braneworld era.

\section{Conclusion}

A natural way to obtain a flat inflaton potential is to invoke some approximate
shift symmetry, which involves a pseudo Nambu-Goldstone boson or extra
components of gauge fields living in extra dimensions. On the other hand,
modifications to the expansion rate of the universe at earlier times turn out
to be crucial for processes that took place in early universe. In this paper we
have examined the observational constraints on the scenario of natural
braneworld inflation, assuming that inflation occurs during the high-energy
regime of the theory. In this framework, the scale at which the global symmetry
is spontaneously broken can be easily lowered to values far below the Planck
mass $M_P$, thus protecting the theory from dangerous nonrenormalizable
operators which could spoil the flatness of the potential.

The explanation of the baryon asymmetry of the universe is another unresolved
issue. Here we have considered the possibility that this asymmetry is generated
by the decay of the inflaton during the reheating era. If the inflaton couples
derivatively to a baryon current, a net asymmetry can be generated as the field
coherently oscillates about its minimum and decays into ordinary matter. In
this case, the spontaneous symmetry breaking scale $f$ also describes the
physics of baryon number violation. In this regard, there may be additional
constraints on the implementation of the mechanism, such as those arising from
proton stability. Finally, if the baryon creation occurs in the braneworld era,
then sufficiently low reheating temperatures and low values of the fundamental
five-dimensional Planck mass $M_5$ can yield the right magnitude of the
observed baryon-to-entropy ratio.

\medskip
\textbf{Acknowledgements}
\medskip

I would like to thank M.~C.~Bento and G.~C.~Branco for reading the manuscript.
This work has been supported by Funda\c c\~ao para a Ci\^encia e a Tecnologia
(FCT, Portugal) under the Grant No.~SFRH/BPD/1549/2000.

\end{document}